\documentclass[prd,aps,nofootinbib,twocolumn,showpacs,showkeys]{revtex4}
\usepackage{amsmath}
\usepackage{amssymb}
\usepackage{epsfig}
\usepackage{graphicx}
\usepackage{axodraw}
\usepackage{color}

\newcommand{\nn}{\nonumber}
\newcommand{\beq}{\begin{equation}}
\newcommand{\eeq}{\end{equation}}
\newcommand{\bea}{\begin{eqnarray}}
\newcommand{\eea}{\end{eqnarray}}
\newcommand{\ba}{\begin{array}}
\newcommand{\ea}{\end{array}}
\newcommand{\bec}{\begin{center}}
\newcommand{\eec}{\end{center}}
\newcommand{\bei}{\begin{itemize}}
\newcommand{\eei}{\end{itemize}}

\def\10{$SO(10)$}
\def\21{SU(2) $\otimes$ U(1) }

\def\422{$SU(4) \otimes SU(2) \otimes SU(2)$}
\def\321{SU(3) $\otimes$ SU(2) $\otimes$ U(1)}

\newcommand {\ignore}[1]{}

\def\lsim{\raise0.3ex\hbox{$\;<$\kern-0.75em\raise-1.1ex\hbox{$\sim\;$}}}
\def\gsim{\raise0.3ex\hbox{$\;>$\kern-0.75em\raise-1.1ex\hbox{$\sim\;$}}}

\newcommand{\AddrAHEP}{%
  AHEP Group, Institut de F\'{\i}sica Corpuscular --
  C.S.I.C./Universitat de Val{\`e}ncia \\
  Edificio Institutos de Paterna, Apt 22085, E--46071 Valencia, Spain}

 \renewcommand{\baselinestretch}{1.0}
\relax
\def\321{$SU(3)\times SU(2)\times U(1)$}

\begin{document}
\preprint{FTUAM 09/16}
\preprint{IFT-UAM/CSIC-09-30}
\preprint{IFIC/09-20}
\renewcommand{\Huge}{\Large}
\renewcommand{\LARGE}{\Large}
\renewcommand{\Large}{\large}
\title{Calculable inverse-seesaw neutrino masses in supersymmetry}
\author{F. Bazzocchi} \email{fbazzoc@few.vu.nl}
\affiliation{Department of Physics and Astronomy, Vrije Universiteit
  Amsterdam, 1081 HV Amsterdam, The Netherlands}
\author{D.~G. Cerde\~no} \email{davidg.cerdeno@uam.es}
\author{C. Mu\~noz} 
\email{carlos.munnoz@uam.es} \affiliation{Departamento de F\'{\i}sica
  Te\'{o}rica C-XI, and 
  Instituto de F\'{\i}sica Te\'{o}rica
  UAM-CSIC, 
  Universidad Aut\'{o}noma de Madrid, 
  Cantoblanco, E-28049
  Madrid, Spain}
\author{J.~W.~F. Valle} \email{valle@ific.uv.es}
\affiliation{\AddrAHEP}

\date{\today}
\begin{abstract}

  We provide a scenario where naturally small and calculable neutrino
  masses arise from a supersymmetry breaking
  renormalization-group-induced vacuum expectation value.  The
  construction consists of an extended version of the Next-to-MSSM and
  the mechanism is illustrated for a universal choice of the soft
  supersymmetry-breaking parameters.  The lightest supersymmetric
  particle can be an isosinglet scalar neutrino state, potentially
  viable as WIMP dark matter through its Higgs new boson coupling.
  The scenario leads to a plethora of new phenomenological
  implications at accelerators including the Large Hadron Collider.

\end{abstract}


\maketitle

Theory has no clue as to what causes the smallness of neutrino
masses. It has become popular to ascribe it to the existence of a very
high scale within the so-called minimal type-I
seesaw~\cite{Minkowski:1977sc}.
Although this approach would fit naturally in unified schemes, no one
to date has produced a convincing unified theory of flavor, where the
observed pattern of quark and lepton masses and mixings is explained,
especially the disparity between the small quark mixing angles and
mixing angles~\cite{Maltoni:2004ei} indicated by neutrino oscillation
experiments.
Moreover, if type-I seesaw is nature's way to understanding neutrinos
one should give up hopes of ever obtaining its direct confirmation by
accelerator experiments, such as the upcoming Large Hadron Collider
(LHC).

Here we adopt as an alternative approach an \321 inverse seesaw
mechanism~\cite{mohapatra:1986bd,gonzalez-garcia:1989rw}, which avoids
introducing new states above the TeV scale.
Neutrino masses arise well below the weak scale, thanks to a very
small singlet mass term in whose presence lepton number is violated.
Naturalness follows in t'Hooft's sense~\cite{'tHooft:1979bh}, namely,
one is allowed to assume the smallness of parameters in whose absence
the symmetry of the theory increases.
Even though this is a perfectly valid and consistent procedure, it has
not become as popular as the high-scale seesaw due to some discomfort
in assuming by hand the smallness of an \321 invariant mass term.
Rather than arguing that such theoretical prejudice is unjustified,
here we provide a plausible mechanism where the origin of such small
scale would, in addition, find a natural dynamical explanation.

Our basic assumption is supersymmetry, the leading framework to
account for the stability of the weak interaction
scale~\cite{Martin:1997ns}.
Here we show how the breaking of supersymmetry can spontaneously
induce the radiative generation neutrino masses at very low scales.
The mechanism requires the existence of a singlet sector, perhaps of
stringy origin~\cite{Witten:1985xc}.  Such sector would be secluded
from the Standard Model sector and hence hardly evolve under the
renormalization group.  ``Calculable'' neutrino masses then arise via
the inverse seesaw mechanism with dynamically generated mass
parameters, in a scenario which can be considered an extended version
of the Next-to-Minimal Supersymmetric Standard Model.

In order to generate naturally small neutrino masses in our scheme we
need to assume the vanishing of some of the new soft-trilinear
parameters. For simplicity, we will illustrate this mechanism imposing
universal conditions for the soft parameters, in a way analogous to
the Constrained next-to-minimal supersymmetric standard model (CNMSSM)
\cite{Hugonie:2007vd}.  However, as we will later clarify, our
dynamical neutrino mass generation scenario need not rely on this
universality.
These initial conditions lead to a consistent phenomenological picture
with an adequate electrically neutral dark matter candidate which is
either be a scalar neutrino, or a spin 1/2 neutralino, with suppressed
couplings.
Of these, here we focus on the first possibility. It has been shown
that in this case the lightest superparticle (LSP) is likely to be a
scalar neutrino whose relic abundance covers the range indicated by
WMAP~\cite{Komatsu:2008hk}, and whose detection cross sections in
nuclear recoil experiments can also be sizeable~\cite{Arina:2008bb}.
Moreover, the required magnitude of the supersymmetric Higgs mass
parameter arises from the expectation value of the extra singlet field
present in the NMSSM~\cite{ellwanger:1993xa}, avoiding the so-called
$\mu$ problem~\cite{kim1984mupa,giudice:1988yz}, as recently advocated
in Ref.~\cite{Cerdeno:2008ep}.

\begin{table}[t]
\begin{center}
\begin{tabular}{|c||ccccccc||ccc|cc|}
  \hline
 & $\hat{Q}$&$\hat{U^c}$&$\hat{D^c}$&$\hat{L}$&$\hat{E^c}$&$\hat{\nu^c}$&$\hat{S}$&$\hat{H^u}$&$\hat{H^d}$&$\hat{\Phi}$&$\hat{\Delta}$ &$\hat{\tilde{\Delta}}$\\
  \hline
  $SU(2)$&2&1&1&2&1&1&1&2&2&1&1&1\\
  $L$&0&0&0&1&-1&-1&1&0&0&0&-2& 1\\
  $R$ &1&1&1&1&1&1&1&0&0&0&0&0 \\
  $Z_3$ &$\omega$&$\omega$&$\omega$&$\omega$&$\omega$&$\omega$&$\omega$&$\omega$&$\omega$&$\omega$&$\omega$&$\omega$\\
  \hline
\end{tabular}
\end{center}
\begin{center}
\begin{minipage}[t]{0.48\textwidth}
\caption[]{Multiplet content of the model. }
\label{tab:mult}
\end{minipage}
\end{center}
\end{table}

In order to illustrate the idea we consider the model defined by the
supermultiplets given in table \ref{tab:mult}, where $L$ denotes the
global continuous lepton number and $R$ denotes the $R$-charge.  We
preserve the $Z_3$ symmetry of the NMSSM which forbids bilinear
couplings, since we still want to generate an EW-scale $\mu$-term in
the same way as in the NMSSM.
The superpotential is given as
\begin{eqnarray}
\label{superpot}
W&&= y^e_{ij} \hat{L_i} \hat{H^d}\hat{E^c_j} +y^u_{ij} \hat{Q_i}
\hat{H^u}\hat{U_j^c}+ y^d_{ij} \hat{Q_i} \hat{H^d}\hat{D_j^c}\nn\\ 
&& +\lambda_1 \,\hat{\Phi} \,\hat{H^u} \hat{H^d} +
\frac{1}{3!}\lambda_2 \,\hat{\Phi}\hat{\Phi} \hat{\Phi}  \nn\\ 
&& + y^\nu_{ij} \hat{Li} \hat{H^u} \hat{\nu^c_j} +\eta_{ij} \hat{\Phi}
\hat{\nu^c_i}\hat{S_j}  \nn\\ 
&&+\frac{1}{2}\xi_{ij}  \hat{\Delta} \hat{S_i}\hat{S_j}
+\frac{1}{2}\tilde{ \xi}   \hat{\Delta}
\hat{\tilde{\Delta}}\hat{\tilde{\Delta}} \ ,
\label{superpotential}
\end{eqnarray}
where the first three terms are standard, the next two account for the
NMSSM extension, and the last four characterize our supersymmetric
inverse seesaw model. In contrast with the simplest versions employed
in Ref.~\cite{Deppisch:2004fa} we have only trilinear
terms, thanks to the $Z_3$ symmetry.
This is also natural in string
constructions, where bilinear couplings are absent from the
superpotential.
The corresponding soft supersymmetry breaking potential reads,
\begin{eqnarray}
\label{soft}
V_{soft}&= a^u_{ij}  \tilde{Q}_i H^u \tilde{u^c}_j + a^d_{ij}  \tilde{Q}_i H^d \tilde{d^c}_j+  a^e _{ij}  \tilde{L}_i H^d \tilde{e^c}_j\nn\\
&+a_{\Phi H} \,\Phi \,H^u\,H^d + \frac{1}{3!}a_\Phi  \Phi^3 + a^\nu_{ij} \tilde{L}_i H^u \tilde{\nu^c}_j\nn\\
&+a^{\eta}_{ij} \Phi \tilde{\nu^c}_i \ S_j + \frac{1}{2} a^S_{ij} \Delta \tilde{S}_i\tilde{S}_j + \frac{1}{2} a_\Delta  \Delta \tilde{\Delta}^2+H.c.\nn\\
& + \Sigma_i m_i^2 |\varphi_i|^2 +\frac{1}{2}\Sigma_i M_i \lambda_i \lambda_i
\end{eqnarray}
where the last two terms are the standard scalar and gaugino soft
supersymmetry breaking terms.

The minimization of the scalar
potential leaves five equations which must be fulfilled for a
successful radiative electroweak symmetry-breaking. Of these, three
are those obtained in the NMSSM and can be used to fix the values of
the vev of the singlet field $\langle\Phi\rangle$, the coupling
$\lambda_2$ and the mass-squared parameter $m_\Phi^2$, a prescription
which is usually adopted in the CNMSSM.  The remaining two equations
relate the vevs of $\Delta$ and $\tilde\Delta$ to the soft parameters
in the secluded sector. These are given as
\begin{eqnarray}
  \label{eq:cond-min}
  \tilde{\xi}^2 v_{\tilde{\Delta}}^3/4 +  \tilde{\xi}^2
  v_{\tilde{\Delta}} v_\Delta^2/2+  a_{{\Delta}} v_{\tilde{\Delta}}
  v_\Delta /\sqrt{2}+m^2_{\tilde{\Delta}}
  v_{\tilde{\Delta}}&=&0\nn\\  
  \tilde{\xi}^2 v_{\tilde{\Delta}}^2 v_\Delta/2 +a_\Delta
  v_{\tilde{\Delta}}^2/(2 \sqrt{2}) +m_\Delta^2  v_\Delta&=&0\,.  
\end{eqnarray}
Due to their different charges with respect to the global lepton
number symmetry, $\Delta$ and $\tilde{\Delta}$ behave differently.  On
the one hand, from the second equation $v_\Delta$ we have
\begin{equation}
  \label{vDelta}
  v_\Delta = -\frac{1}{\sqrt{2}}a_\Delta
  \frac{v_{\tilde{\Delta}}^2}{\tilde{\xi}^2 v_{\tilde{\Delta}}^2+2
    m_\Delta^2}\,. 
\end{equation}
The light neutrino mass is controlled by $v_\Delta$ through the
inverse seesaw
mechanism~\cite{mohapatra:1986bd,gonzalez-garcia:1989rw} which applied
to the lagrangian obtained by the superpotential of
eq.\,(\ref{superpot}) gives
\begin{equation}
  \label{massnu}
  m^\nu=-  y^\nu v^u \, (\eta v_\Phi)^{-1}\, (\xi v_\Delta)  ((\eta
  v_\Phi)^T)^{-1} ( y^\nu v^u )^T\,. 
\end{equation}
From eq.\,(\ref{vDelta}) we see that $v_\Delta \propto a_\Delta$, so
it depends exclusively on the trilinear soft parameter of the singlet
sector. 
On the other hand, the equation involving $v_{\tilde{\Delta}}$ gives in
first approximation 
\begin{equation}
  v_{\tilde{\Delta}}^2=-4\frac{m_{\tilde{\Delta}}^2}{
    \tilde{\xi}^2}+O( a_\Delta)\,. 
\end{equation}

This is controlled by the soft parameter $m_{\tilde{\Delta}}^2$ and
does not vanish in the limit in which the trilinear soft term
$a_\Delta$ goes to zero. Thus under the assumption that the trilinear
soft terms of the singlet sector vanish at the unification scale,
minimizing the tree-level Higgs potential one finds that
$\tilde{\Delta}$ develops a vacuum expectation value
$v_{\tilde{\Delta}}$ while ${\Delta}$ does not. Therefore, in this
tree level limit neutrinos are still massless.

\begin{figure}[!t]
\begin{center}
\includegraphics[angle=0,height=3cm]{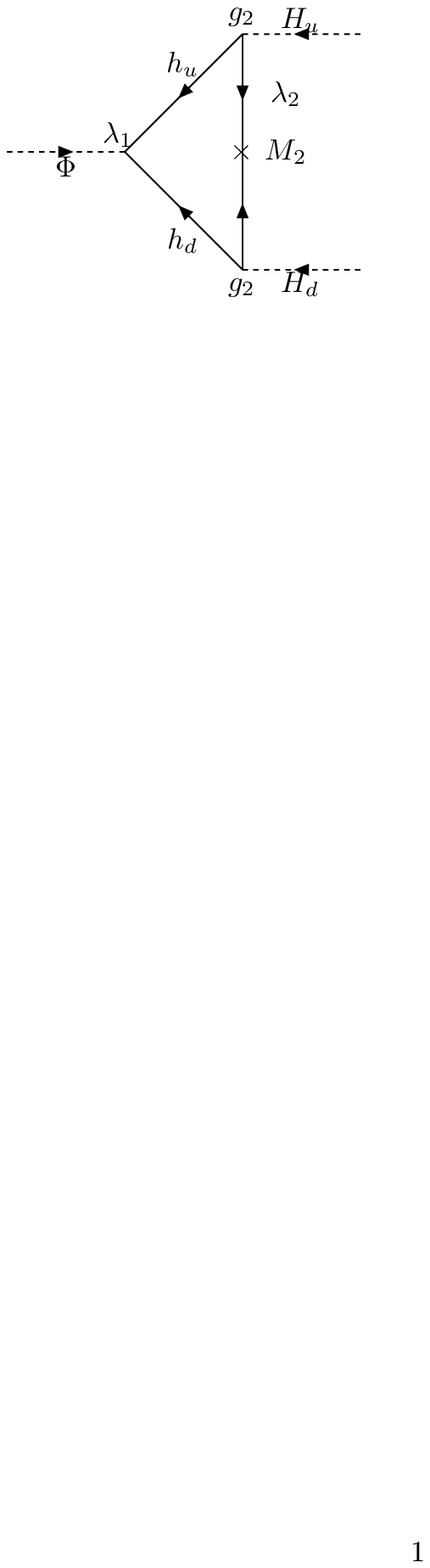}
\caption{Supersymmetry breaking as seed for low-scale dynamical
  neutrino mass generation.}
\label{fig:aseed} 
\end{center}
\end{figure}

One can derive the corresponding logarithm renormalization group
evolution equations (RGEs) for all masses and couplings. These contain
the RGEs of the NMSSM, supplemented with the evolution of the new
parameters of the secluded sector. Imposing vanishing 
trilinear
soft terms at the unification scale, the
evolution of the corresponding 
trilinear soft parameters can be approximated  by the following set
of equations 
\begin{eqnarray}
 \beta_{a_{\Phi H}}&=&6\lambda_1   (g_2^2 M_2+ g_1^2 M_1/5)\,,\nn\\ 
 \beta_{a_\Phi}&=& 12 a_{\Phi H}   \lambda_1^* \lambda_2   \,, \nn\\ 
 \beta_{a_{\eta}}&=&4 a_{\Phi H} \lambda_1^*  \eta\,, \nn\\ 
 \beta_{a_S}&=& 2 a_{\eta} \eta^* \xi \,, \nn\\ 
 \beta_{a_\Delta}&=&  a_S \xi^* \tilde{\xi}\,,
 \label{RGEs}
\end{eqnarray}
where we have defined $16\pi^2\, da_i/dt=\beta_{a_i}$.  The
first line above encodes the fact that gaugino soft supersymmetry
breaking terms are the seed for 
the low-scale dynamical neutrino mass generation
mechanism proposed here, as illustrated in Fig.~\ref{fig:aseed}.

The breaking of supersymmetry due to the non-vanishing gaugino mass is
sequentially transmitted to all the soft trilinear terms of the
singlet sector, the smallest being $a_{\Delta}$. Their approximated
order of magnitude can be easily read from the RGEs given in
eq.\,(\ref{RGEs}) and are given by 
\begin{eqnarray} 
  \label{eq:ad2}
  a_{\Phi H}\sim M_{1/2}  O(g^2) 
  \left(\frac{\log \mu/M_{0}}{16 \pi^2}\right) \,, \nn\\
  a_{\Phi}~,~a_{\eta}~\sim  M_{1/2}  O(g^2) 
  \left(\frac{\log \mu/M_{0}}{16 \pi^2}\right)^2 \,, \nn\\  
  a_{S}\sim M_{1/2}  O(g^2) 
  \left(\frac{\log \mu/M_{0}}{16 \pi^2}\right)^3  \,, \nn\\  
  a_{\Delta} \sim   M_{1/2}  O(g^2) 
  \left(\frac{\log \mu/M_{0}}{16 \pi^2}\right)^{4}\,.
\end{eqnarray}
The presence of a small but non vanishing $a_\Delta$, which arises
effectively at four-loops\footnote{Because of the singlet nature of
  $\Delta$ and $\tilde\Delta$ and the specific couplings in
  eq.\,(\ref{superpot}) 
  the RGE two-loop order contributions to $a_\Delta$ are absent. We
  did not explicitly check that all the three-loop order ones vanish,
  even if we expect so. However, even if the latter were induced, one
  could arrange the arbitrary tree level parameters  
  to obtain $v_\Delta$ of the correct order of magnitude.},  
induces in turn a naturally small vev
$v_{\Delta}$ according to eq.\,(\ref{vDelta}).  As a result $v_\Delta$
is naturally expected to be very small, of order MeV or even smaller.
This shows how the small calculable parameter $v_\Delta$ acts as the 
seed of the light neutrino mass according to eq.\,(\ref{massnu}). The
existence of the secluded sector therefore plays an important role in
our proposal of a novel mechanism in terms of which to understand the
origin of neutrino mass.
As we have already explained, the main assumption is that the
trilinear parameters associated to the singlet sector, $a_{\Phi H}$,
$a_{\Phi}$, $a_{\eta}$, and $a_{S}$ vanish at the unification scale
$M_0$, while the rest of the soft parameters are unconstrained.
Vanishing trilinear parameters are possible in the supergravity
scenarios obtained as the low-energy limit of string theory. More
specifically in some phenomenologically appealing constructions, such
as D-brane compactifications of the Type-I string or orbifold
scenarios in the Heterotic string, some trilinear couplings vanish
for specific choices of the parameters which define supergravity
breaking (see e.g.  some of the examples in
Refs.\,\cite{Cerdeno:2001se}). 
Notice that the fields in the new singlet sector in our
model have different quantum numbers than the MSSM fields and must
therefore correspond to different string modes. Thus the
vanishing of the trilinear terms associated to these singlets does not
necessarily imply that all the trilinears are null.

Let us now illustrate this mechanism with a specific example.  Since
the role of the MSSM trilinear parameters is not relevant for our
discussion, for simplicity we will impose universal conditions just as
in the CNMSSM and assume that all trilinear soft breaking terms vanish
at $M_0$. We also consider a universal gaugino mass parameter,
$m_{1/2}$, and a positive scalar squared-mass parameter, $m^2_0$, for
all gauge non-singlet scalars.  For the singlet scalars we assume them
to be all equal to $m_\Phi^2$ and use the minimization conditions to
fix its value, which typically results in $m_\Phi^2<0$. This choice
could be understood, e.g., if the singlet fields had a different
origin from the NMSSM fields in a more fundamental theory. Negative
squared-mass parameters for scalars were also considered in the MSSM
\cite{Feng:2005ba}. 

\begin{figure}[!t]
\begin{center}
\includegraphics[angle=0,width=0.48\textwidth]{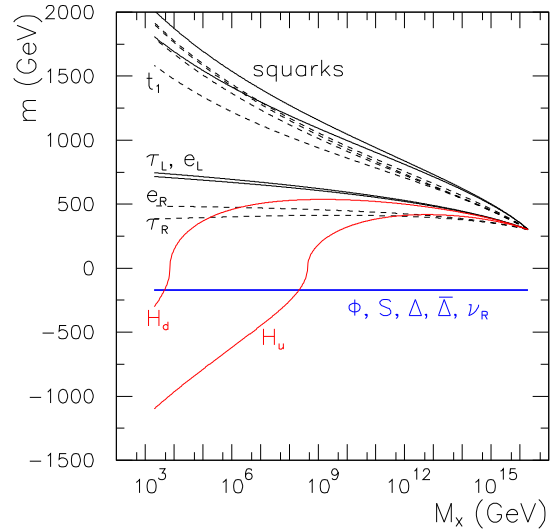}
\vglue -.3cm
\caption{ Renormalization group evolution of the scalar
    masses (with $m\equiv {\rm sign}(m_o)\sqrt{|m_0^2|}$) 
  for a representative choice of parameters (see text). }
\label{fig:run}
\end{center}
\end{figure}
In Fig.\,\ref{fig:run} we display the evolution of the masses of the
scalar fields of the model for a concrete example in which we choose
$m_{1/2}=1$~TeV, $m_o=300$~GeV, $\tan\beta=40$, $\lambda_1=0.01$,
$\xi=0.1$ and $\eta=\tilde\xi=0.0053$.  The minimization conditions
impose $\mu=1140$~GeV, $m_\Phi^2=-2.94\times10^4$~GeV$^2$ and
$\lambda_2=-0.0011$ at the EW scale \footnote{ As noted in
    \cite{Djouadi:2008uj}, the resulting value of $\lambda_2$ after
    solving the minimization conditions in the CNMSSM is typically one
    order of magnitude smaller than $\lambda_1$.}.  We note that the
evolution of the parameters concerning the states in the NMSSM are not
substantially different from what is expected and that, due to the
smallness of $\lambda_2$, the soft masses of the singlets practically
do not deviate from their value at $M_0$.

After numerically solving the RGEs, the particle spectrum can be
calculated at the EW scale. The inclusion of the secluded sector has
no effect on the masses of most of the NMSSM particles. However, the
new singlet $S$ mixes with the right- and left-handed sneutrino
states, giving rise to three sneutrino states.

One can see that in most cases the lightest sneutrino is a combination
of the two singlet states $\tilde{\nu}^c$ and $\tilde{s}$. Since all
trilinear couplings vanish at the unification scale, the trilinears
involving only gauge singlet fields run very slowly so $v_\Delta$ is
very small compared to the other vevs. Thus for the above choice the
sneutrino mass matrix can be approximated as
\begin{equation}
M^2_{\tilde{\nu}_{ri}}\sim \left( \begin{array}{ccc} m^2_{{L}}  &0 &
  0 \\ 0 &  m^2_0+ \alpha_{ri}  v^2  & \pm  \delta v^2 \\  0& \pm
  \delta  v^2 &   m^2_0 +  \beta_{ri}
  v^2  \end{array}\right)\,.\nonumber 
\end{equation}
with $0<\alpha_{ri},\beta_{ri}\sim \mathcal{O}(1)$ while $ \delta\sim
\mathcal{O}(0.1)$ and where we have used $m_{{\nu}^c}^2,
m_S^2~\tilde{m_\Phi}^2<0$.  Given that $ m^2_L>0$, the lightest
sneutrino is typically the CP even or CP odd combination of the
singlet states.

\begin{figure}[!t]
\begin{center}
\includegraphics[angle=0,height=9.5cm]{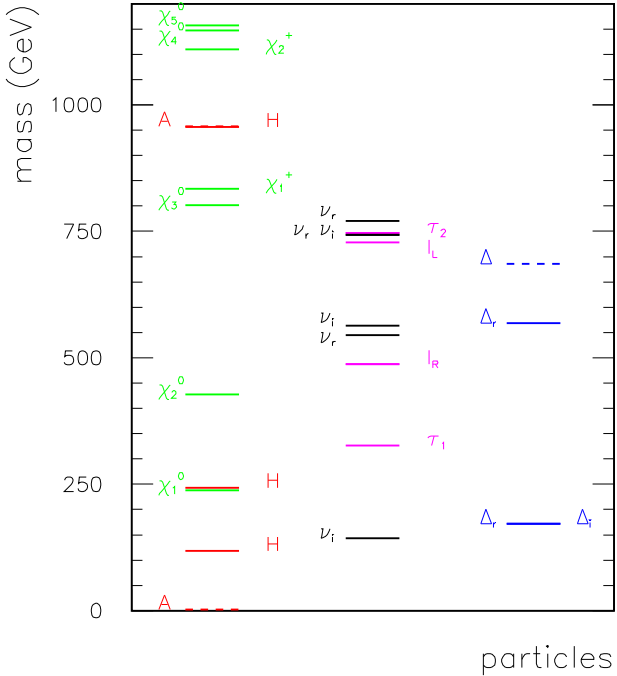}
\vglue -.3cm
\caption{Supersymmetric spectrum for the same choice of
  parameters as in Fig.\,\ref{fig:run}. The real(imaginary) sneutrino
  state is labelled as $\nu_r$($\nu_i$), the scalar(pseudoscalar) of
  the secluded sector is indicated as $\Delta_r$($\Delta_i$)
  and the singlino as $\Delta$. Gluino and squark masses are larger
  than $1.5$~TeV and not shown.
}
\label{fig:spec}
\end{center}
\end{figure} 

As illustrated in Fig.~\ref{fig:spec}, due to these mixing effects it
is likely that the lightest supersymmetric particle is a mainly a
mixture of the singlet scalars in $\nu^c$ and $S$ , instead of the
neutralino or stau. This is analogous to the construction of
Ref.\cite{Arina:2008bb}, although now the sneutrino has virtually no
left-handed component.
In this sense, this model is similar to the scenario of
Refs.~\cite{Cerdeno:2008ep,Cerdeno:2009dv}, since in both the
right-handed sneutrino component couples directly to the NMSSM Higgs
sector through the singlet $\Phi$. As shown there, this makes it
possible to fulfil the WMAP result, thereby making the sneutrino a
viable WIMP. A similar effect is expected in the present model, this
time through the $\eta$ coupling in eq.\,(\ref{superpotential}).
Thus, the scheme proposed here opens yet new alternative ways to
understand supersymmetric dark matter.

In this  example, the smallness of the $\lambda_2$ parameter
implies the quasi-restoration of a U(1) Peccei-Quinn symmetry in the
superpotential. This entails the occurrence a very light CP-odd Higgs
(in our example $m_A=3$~GeV) as the pseudo-Goldstone boson of this
broken symmetry \cite{Dobrescu:2000yn}. Since the latter is almost a
pure singlet it is consistent with existing phenomenological bounds.

\begin{figure}[!t]
\begin{center}
\includegraphics[angle=0,height=9.5cm]{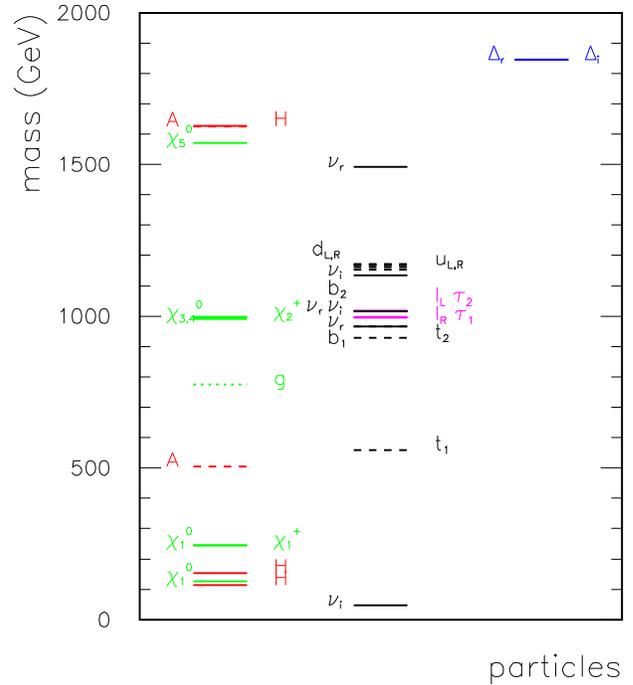}
\vglue -.3cm
\caption{Same as in Fig.\,\ref{fig:spec}, but for a different choice
  of input parameters, with larger $\lambda_1$ and $\lambda_2$, and a
  small $\tan\beta$, as described in the text.  }
\label{fig:spec2}
\end{center}
\end{figure} 

Examples with a larger value of $\lambda_1$ generally lead to a larger
$\lambda_2$ when the REWSB conditions are solved. In these cases the
lightest CP-odd Higgs is not as light as in the example above and,
apart from of some new heavy states, the spectrum resembles that of
the MSSM.  However, the increase in $\lambda_1$ also entails a more
negative value of $m_\Phi^2$ as a solution to the REWSB equations, and
this in turn makes it more complicated to obtain non-tachyonic
sneutrinos. Only through an increase in $\eta$ can this be achieved
but this has as a consequence a significant increase in the masses of
the particles in the secluded sector ($\Delta$ and $\Delta_{r, i}$).
For completeness, in Fig.\,\ref{fig:spec2} we display the resulting
spectrum for $m_{1/2}=300$~GeV, $m_o=1$~TeV, $\tan\beta=2$,
$\lambda_1=1$, $\xi=\tilde\xi=0.07$ and $\eta=2.8$.  The minimization
conditions now lead to $\mu=996$~GeV,
$m_\Phi^2=-1.17\times10^6$~GeV$^2$ and $\lambda_2=-0.4$ at the EW
scale. As in the previous example, the imaginary sneutrino is now the
LSP and this has a significant mixture of the scalars in $\nu^c$ and
$S$. Notice that the particles in the secluded sector are heavier than
$1.7$\,TeV.

Last, but not least, in addition to a new supersymmetric dark matter
scenario, our model leads to different phenomenological implications
for the LHC and other accelerator experiments. 
The most interesting of these 
follow directly or indirectly from the new gauge singlet fermions at
the TeV-scale.
Apart from the possibility of direct production through mixing in the
Standard Model weak currents~\cite{Dittmar:1990yg}, their exchange can
induce lepton-flavor violating (LFV)~\cite{bernabeu:1987gr} as well as
leptonic CP violating effects~\cite{branco:1989bn}, leading to
processes such as $\mu^-\to e^-\gamma$, nuclear $\mu^--e^-$
conversion~\cite{Deppisch:2005zm} and LFV tau
decays~\cite{Ilakovac:1994kj}.
These processes can proceed even in the limit of decoupled
supersymmetry, and even in the absence of neutrino masses. As a result
their expected rates can be sizeable~\cite{Deppisch:2004fa}.
In addition supersymmetry brings in the possibility of observing
lepton flavor violation at high energies, in the decays of
supersymmetric states, opening the possibility that LHC can directly
probe the underlying physics~\cite{Esteves:2009vg}.

Finally 
$v_{\tilde{\Delta}}$
spontaneously breaks lepton number at the TeV scale, generating a
pseudoscalar Goldstone boson, called Majoron~\cite{Chikashige:1980ui}.
Its couplings with ordinary matter are tiny, evading stellar energy
loss constraints. %

In summary, we have described a framework in which supersymmetry
breaking can provide the dynamical origin for small neutrino masses
through the inverse seesaw mechanism.
The seed for neutrino masses is a small renormalization-group-induced
\321 singlet vacuum expectation value, while the $\mu$ problem is also
dynamically solved as in the NMSSM. 
A mixed singlet sneutrino arises as a natural candidate for WIMP dark
matter, in addition to a plethora of new phenomenological
implications. 

We thank O. Seto for participation at the early stages of this work.
This work was supported by the Spanish grants FPA2009-08958, 
HEPHACOS, S-2009/ESP-1473 (Comunidad de
Madrid) and Prometeo/2009/091, and  MICINN's Consolider-Ingenio
2010 Program MULTIDARK CSD2009-00064, and by the EU networks
MRTN-CT-2006-035863, and PITN-GA-2009-237920.
FB was partially supported by the foundation for Fundamental Research
of Matter (FOM) and the National Organization for Scientific Research
(NWO).  DGC was supported by the ``Ram\'on y Cajal'' program of the
MICINN.

\renewcommand{\baselinestretch}{1}

\end{document}